\documentclass{optica-article}

\journal{opticajournal} % for journals or Optica Open

\articletype{Research Article}

\usepackage{lineno}
\usepackage{soul}
\newtheorem{definition}{Definition}

\def\ania{{\hat{a}^\dagger}} 
\def\anib{{\hat{b}^\dagger}}
\def\anic{{\hat{c}^\dagger}}
\def\anid{{\hat{d}^\dagger}}
\def\anie{{\hat{e}^\dagger}}
\def\anif{{\hat{f}^\dagger}}
\def\anig{{\hat{g}^\dagger}}

\def\>{{\rangle}}

\def\lossBC{{\sqrt{1-\eta_{BC}^2}}}
\def\lossSC{{\sqrt{1-\eta_{SC}^2}}}
\def\lossSD{{\sqrt{1-\eta_{SD}^2}}}

\begin{document}
	
\title{Heralded optical entanglement distribution via lossy quantum channels: A comparative study}

\author{Wan Zo,\authormark{1,2} Seungbeom Chin,\authormark{3,4,*} and Yong-Su Kim\authormark{1,5,\dag}}

\address{\authormark{1}Center for Quantum Technology, Korea Institute of Science and Technology (KIST), Seoul, 02792, Republic of Korea\\
	\authormark{2}Department of Physics and Astronomy, Seoul National University, Seoul 08826, Korea\\
	\authormark{3}Department of Electrical and Computer Engineering, Sungkyunkwan University, Suwon 16419, Republic of Korea\\ 
	\authormark{4}Okinawa Institute of Science and Technology Graduate University, Okinawa 904-0495, Japan\\ 
	\authormark{5}Division of Quantum Information, KIST School, Korea University of Science and Technology, Seoul 02792, Republic of Korea}
	
	\email{\authormark{*}sbthesy@gmail.com} %% email address is required; see note below about the corresponding author designation
	\email{\authormark{\dag}yong-su.kim@kist.re.kr} %% email address is required; see note below about the corresponding author designation

	% use {asbstract*} to suppress the copyright line. Copyright information will be added in production

\begin{abstract*} 
Quantum entanglement serves as a foundational resource for various quantum technologies. In optical systems, entanglement  distribution rely on the indistinguishability and spatial overlap of photons. Heralded schemes play a crucial role in ensuring the reliability of entanglement generation by detecting ancillary photons to signal the creation of desired entangled states. However, photon losses in quantum channels remain a significant challenge, limiting the distance and capacity of entanglement distributions. This study suggests three heralded schemes that distribute multipartite Greenberger-Horne-Zeilinger (GHZ) states via lossy quantum channels. These schemes utilize different photon sources (Bell states or single-photons) and channel structures (centralized or decentralized heralding detectors). By comparing success probabilities and heralding efficiency, we find that each scheme has its own advantage according to the number of parties and the channel distance and the security requirement. This analysis provides insights into designing resilient heralded circuits for quantum information processing over lossy channels.
\end{abstract*}
	
	\section{Introduction}
	
\begin{figure}[t]
		\centering
		\includegraphics[width=0.6\textwidth]{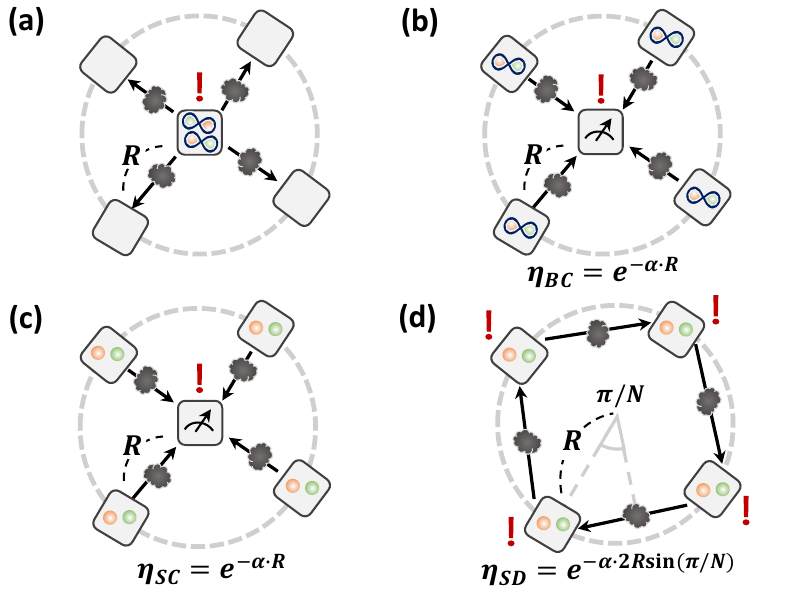}
		\caption{Entanglement distribution schemes via lossy channels. Here, $\eta$ denotes the channel transmission coefficient, $\eta = e^{-\alpha l}$, with respect to the channel length $l$ and the attenuation constant $\alpha$. The channel length $l$ can be represented by the radial size of the quantum networks, $R$. (a) Direct entanglement distribution from a central third-party. (b) Bell state inputs with a centralized third party (BC) scheme. (c) Single-photon inputs with centralized third-party (SC) scheme. (d) Single-photon inputs with decentralized heralding (SD) scheme. While the channel length $l$ of the BC and SC schemes is $l=R$, that of the SD scheme is $l=2R\sin{\pi/N}$ due to network structure.}
%\caption{Entanglement distribution schemes via lossy channel. $\eta$ denotes channel transmission coefficient with respect to channel length $l$. $R$ and $\alpha$ denote the radial size of quantum networks and attenuation coefficient. \textbf{(a)} Direct entanglement distribution from central third-party.  \textbf{(b)} Bell state inputs with centralized third-party (BC) scheme. It is equivalent to N-party entanglement swapping. The channel length $l=R$. \textbf{(c)} Single-photon inputs with centralized third-party (SC) scheme. The channel length $l=R$. \textbf{(d)} Single-photon inputs with decentralized heralding (SD) scheme. The channel length $l=2R\sin{\pi/N}$. Heralding detectors are evenly distributed among $N$ parties.}
		\label{scheme}
	\end{figure}

Entanglement among distant parties is a crucial resource for various quantum applications, such as quantum networks~\cite{ursin2007entanglement,simon2017towards,wehner2018}, quantum cryptography~\cite{ekert1991quantum,yin2020entanglement}, distributed quantum sensing~\cite{liu2021distributed,kim2024distributed}, and quantum computing~\cite{klm2001,barz2012demonstration,fitzsimons2017private,huang2017experimental}. The photonic system is one of the most promising candidates for such applications due to their ability to utilize existing fiber-based networks and the rapid development of integrated photonics. Since entanglement cannot be generated via local operations and classical communication, entangled photons are usually generated at a localized place and distributed to distant parties via optical channels, see Fig.~\ref{scheme}(a). However, optical channels are prone to loss, thus the distant parties can only share entangled photons upon successful transmission that cannot be determined unless quantum non-demolition (QND) measurements are employed~\cite{braginsky1996quantum, grangier1998quantum}. Since practical QND measurements in optical systems are still beyond with current technology, entanglement distribution via direct transmission relies on post-selection.
	
To achieve scalable quantum information processing in a large quantum network, entanglement distribution should be heralded by ancillary photons~\cite{walther2007heralded, wagenknecht2010, caspar2020heralded, gubarev2020improved}. For instance, the standard entanglement swapping protocol shown in Fig.~\ref{scheme}(b) can be understood as a heralded distribution of entanglement among distant parties~\cite{pan1998experimental, pan1998greenberger, bose1998multiparticle}. The successful entanglement measurement at the central party occurs only after all photons from the distant parties have successfully reached it. However, the standard entanglement swapping protocol relies on synchronized initial entangled states among distant parties that are challenging to implement. Therefore, single-photon inputs instead of initial entanglement inputs would offer a more manageable alternative for heralded schemes.

	%%%%%%%%%%%%%%%%%%%%%%%%%%%%%%%%%%%%%%
	
	 In this work, we propose and analyze three different types of heralded schemes for distributing multipartite entanglement among distant parties, with a focus on the effects of channel loss. In particular, we study distribution of the Greenberger-Horne-Zeilinger (GHZ) states which is widely used as a resource for various quantum information processes~\cite{bouwmeester1999, brunner2014bell, li2015resource}. The three schemes are distinguished by their photon sources and channel structures. The first scheme, dubbed the BC scheme (Bell state inputs with a Centralized third party, see Fig.~\ref{scheme}(b)), is the standard entanglement swapping protocol where Bell states are used as photon sources and entanglement measurements are performed at a central third party. While this scheme possess advantages on the success probability and heralding efficiency, it is challenging to simultaneously generate Bell states among distant parties. Therefore, we investigate the second scheme, called the SC scheme (Single-photon inputs with a Centralized third party, see Fig.~\ref{scheme}(c)), which replaces the initial Bell states with single photons while still employing a central third party to perform entanglement measurements. This scheme can be understood as an expansion to multi-partite scenario of Ref.~\cite{zhang2008,lasota2014linear}. Lastly, we explore a heralded scheme that uses single photon sources without a centralized third party (hence called decentralized, SD scheme. See Fig.~\ref{scheme}(d))~\cite{chin2024,chin2023}. 
  %On the other hand, a heralded scheme with single photon sources that does not include the intervention of a third party to herald the entanglement (hence called \emph{decentralized}) is recently introduced in Ref.~\cite{chin_w}. 
 %\purple{The advantage of the SD scheme over the BC and SC schemes is that its heralding information is equally shared among each party that enhances the security of quantum communications.}

  %\purple{We will compare the robustness of these three schemes under lossy channels based on the criteria of success probabilities ($P_{suc}$) and heralding efficiencies ($H_{eff}$, the definitions of $P_{suc}$ and $H_{eff}$ are given in Section~\ref{criteria}). Our analysis reveals that each scheme has its own advantage depending on the number of parties and the distance among parties.}

The remainder of this paper is organized as follows: In Section~\ref{evolution}, we first propose the criteria, i.e., the success probability and the heralding efficiency, to compare the performance of the three heralded schemes. Then we introduce three heralded schemes (BC, SC, and SD) for generating the GHZ state and calculate the success probability and heralding efficiency for each scheme with channel losses. In Section~\ref{comparative_study}, we evaluate the utilities of the schemes based on the criteria of network structures, success probabilities, and heralding efficiencies. In Section~\ref{conclusion}, we summarize our results and discuss possible future research topics.

\section{Entanglement distribution via lossy channels}  \label{evolution}

\subsection{The criteria}\label{criteria}

We employ two criteria, the success probabilities $P_{suc}$ and heralding efficiency $H_{eff}$, to compare the above schemes with lossy channels. The success probability $P_{suc}$ is the probability of generating the target state, i.e., the GHZ state, among all trials of transmitting the initial states via channels. The heralding efficiency $H_{eff}$ is the GHZ state generation probability upon with the successful heralding detections. These two values are defined more rigorously as follows:

%%%%%%%%%%%%%%%%%%
\begin{definition}
	The success probability $P_{suc}$ is the probability that participating parties share target entangled state among all trials.
\end{definition}
%%%%%%%%%%%%%%%%%%

%%%%%%%%%%%%%%%%%%
\begin{definition}
	The heralding probability $P_{hr}$ is the probability that the heralding detection provides a signal in the way we have designed among all trials. Ideally, heralding detection indicates successful distribution of entangled state.
\end{definition}
%%%%%%%%%%%%%%%%%%

%%%%%%%%%%%%%%%%%%
\begin{definition}
	The heralding efficiency $H_{eff}$ is given as the conditional probability with $P_{suc}$ and $P_{hr}$ by
	\begin{align}
		H_{eff} = \frac{P_{suc}}{P_{hr}}. 
	\end{align}
\end{definition}
%%%%%%%%%%%%%%%%%%
%In the following, it will be shown that $P_{suc}$ and $P_{hr}$ can be  less than 1 even when optical channels are lossless. 

%\teal{it is possible to create schemes with $H_{eff}=1$ for a specific input in the lossless channels. The three schemes discussed in the text also have $H_{eff}=1$ when there is no loss.} When the channels become lossy, however, $H_{eff}$ can be less than 1.  

\noindent Now, we investigate three different heralded optical schemes, i.e., BC, SC, and SD schemes, for generating $N$-partite GHZ states. We provide step-by-step state evolution and calculate the success probabilities $P_{suc}$ and heralding efficiencies $H_{eff}$ in the presence of channel losses. %A comparative analysis on the success probabilities and heralded efficiencies of the three schemes will be provided in Section~\ref{comparative_study}. 

\subsection{Bell state inputs with centralized third-party (BC scheme)}\label{centr_bell}

\begin{figure}[t]
	\centering
	\includegraphics[width=0.75\textwidth]{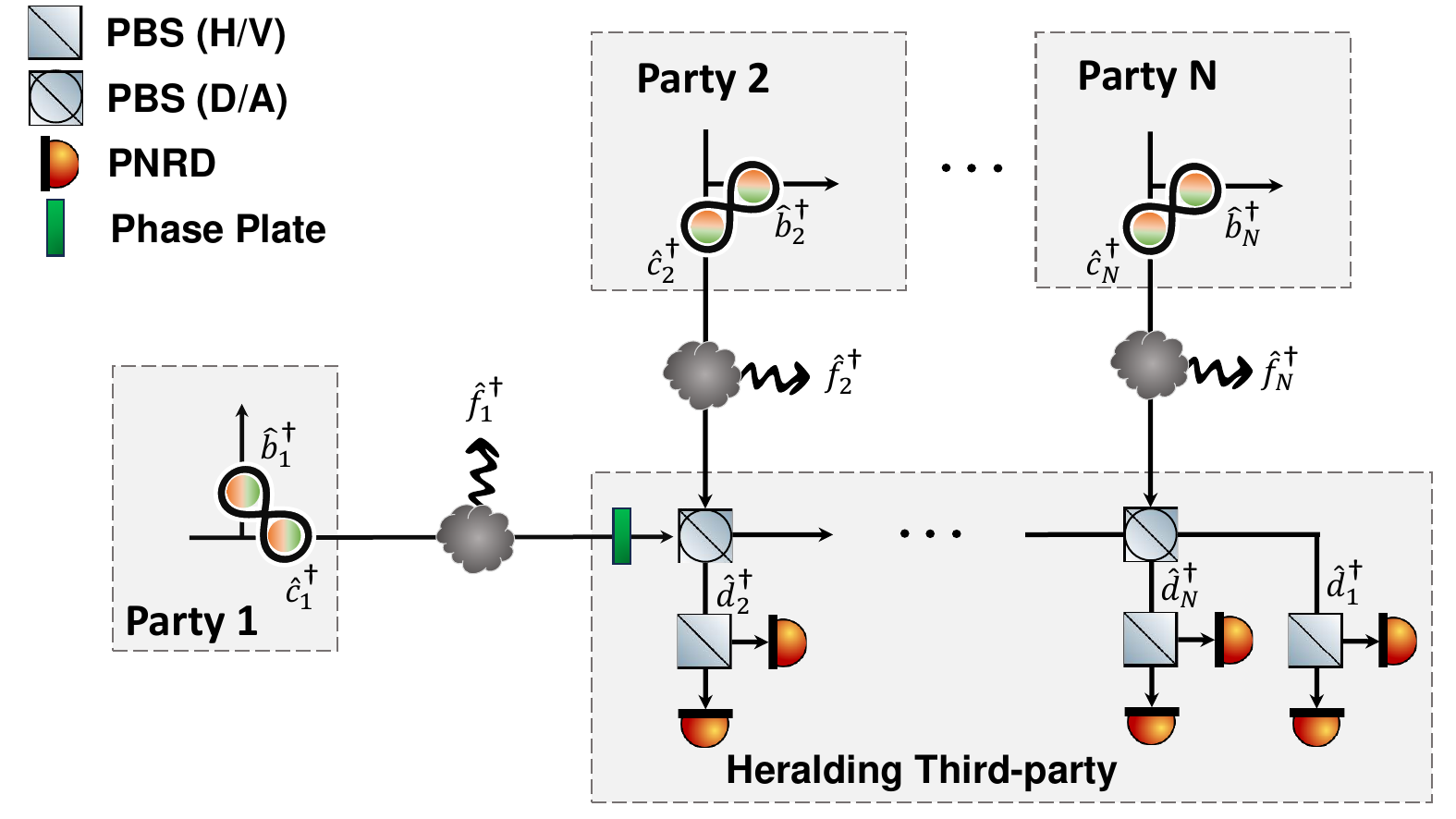}
	\caption{Schematic of the BC heralding scheme. The setup consists of linear optics and photon-number resolving detectors (PNRDs). Input resources are $N$ Bell states, resulting in an $N$-party GHZ state}
	\label{fig_bell_central}
\end{figure}

We can generate the $N$-partite GHZ state with $N$ sets of Bell states (thus, $2N$ photons) as described in Fig.~\ref{fig_bell_central}, which can be considered as a $N$-partite generalization of the entanglement swapping~\cite{pan1998experimental}. Initially, each party labelled $i$ $\in \{1,2,\cdots, N\}$ prepares for the Bell state
\begin{align}
	|\psi^+\>_i
	%= \frac{|HV\>_i +|VH\>_i}{\sqrt{2} } 
	=\frac{1}{\sqrt{2}} (\anib_{i,H}\anic_{i,V} + \anib_{i,V}\anic_{i,H})|vac\>,
\end{align} where the subscripts $H$ and $V$ represent horizontal and vertical polarization states, respectively.
Then the overall $2N$-photon initial state $|\psi\>_{in}$ is given as
\begin{equation}
	\begin{aligned}
		|\psi\> = \prod_{i=1}^N |\psi^+\rangle_i = \left(\frac{1}{\sqrt{2}}\right)^N \prod_{i=1}^{N} (\anib_{i,H} \anic_{i,V} + \anib_{i,V} \anic_{i,H}) |vac\rangle.
		\label{cnt_bell_1}
	\end{aligned}
\end{equation}
Each party retains one of the entangled photons  ($\anib$) and sends the other ($\anic$) to the central third-party.

%%%%%%%%%%%%%%%%%%%%%%%%%%%%%%%%%%%%%%
%%%%%%%%%%%%%%%%%%%%%%%%%%%%%%%%%%%%%%

Photons forwarded to the central third-party can be lost through channels, which is represented as follows:	
\begin{equation}
	\begin{aligned}
		&\anic_{i}\rightarrow \eta_{BC} \anic_{i} +\sqrt{1-\eta_{BC}^2}\anif_i,
		\label{cnt_bell_2}
	\end{aligned}
\end{equation}
where $\eta_{BC}$ denotes the transmission coefficient of the BC scheme and $\hat{f}_i^\dag$ is the environment mode corresponding to the loss. As drawn in the heralding third-party of Fig.~\ref{fig_bell_central}, the photons sent to the third-party encounter the polarization beam splitter (PBS), which splits diagonal ($|D\> \equiv (|H\>+|V\>)/ \sqrt{2}$) and anti-diagonal ($|A\> \equiv (|H\>-|V\>)/ \sqrt{2}$) polarization states. Thus, the transform is:
\begin{equation}
	\begin{aligned}
		\anic_{i,H}\rightarrow \frac{1}{\sqrt{2}} \left(\anid_{i,D} + \anid_{i\oplus 1,A} \right),~~~\anic_{i,V}\rightarrow \frac{1}{\sqrt{2}} \left(\anid_{i,D} - \anid_{i\oplus 1,A} \right),
		\label{cnt_bell_3}
	\end{aligned}
\end{equation}
where $\oplus$ denotes a summation modulo $N$. Note that we compensate phase shift $e^{i\pi}$ for path mode $\anic_1$ with a phase plate. As will be revealed later, the relative phase here only affects the relative phase of the generated GHZ state and does not affect the success probability or the heralding efficiency.

The wave function is written as
\begin{equation}
	\begin{aligned}
		&|\psi\> \rightarrow \left(\frac{1}{\sqrt{2}}\right)^N \prod_{i=1}^N \Bigg [ \frac{\eta_{BC}}{\sqrt{2}} \left(\anib_{i,H}+\anib_{i,V} \right) \anid_{i,D} +\frac{\eta_{BC}}{\sqrt{2}}\left(\anib_{i,H}-\anib_{i,V} \right)\anid_{i\oplus 1,A}\\ &+\lossBC(\anib_{i,H}\anif_{i,V}+\anib_{i,V}\anif_{i,H}) \Bigg]|vac\>.
		\label{cnt_bell_5}
	\end{aligned}
\end{equation}
We postselect only the cases when one photon arrives at each photon-number resolving detector (PNRD), which {\it heralds} the target GHZ state. Hence the terms that contribute to the final state will contain $\prod_{i=1}^{N}\anid_i$. We can eliminate other path modes in the state evolution, thus the quantum state is:
%%%%%%%%%%%%%%%%%%%%
\begin{equation}
	\begin{aligned}
		|\psi\> \rightarrow \left(\frac{\eta_{BC}}{\sqrt{2}}\right)^N\Bigg[\prod_{i=1}^N \anib_{i,D} \anid_{i,D} + \prod_{j=1}^N \anib_{j,A} \anid_{j,A}\Bigg]|vac\>.
		\label{cnt_bell_6}
	\end{aligned}
\end{equation}
%%%%%%%%%%%%%%%%%%%%
Equation~\eqref{cnt_bell_6} indicates the quantum state corresponding to heralding signal from PNRD. 
Note that Eq.~\eqref{cnt_bell_6} is written in D/A polarization basis, but measurement basis in the heralding detection mode $\anid$ is H/V. Thus, the full expansion of the equation is as follows:
\begin{equation}
	\begin{aligned}
		&|\psi\> \rightarrow \frac{\eta_{BC}^N}{\sqrt{2}^{2N-1}}\bigg[ \left(  \frac{\anib_{1,D} ... \anib_{N,D} + \anib_{1,A} ... \anib_{N,A}}{\sqrt{2}}\right) \anid_{1,H}...\anid_{N,H} \\
		&+\left( \frac{\anib_{1,D} ... \anib_{N,D} - \anib_{1,A} ... \anib_{N,A}}{\sqrt{2}}\right) \anid_{1,H}...\anid_{N,V} \\
		&+ ...\\
		&+ \left( \frac{\anib_{1,D} ... \anib_{N,D} + (-1)^{N}\anib_{1,A} ... \anib_{N,A}}{\sqrt{2}}\right) \anid_{1,V}...\anid_{N,V} \bigg]|vac\>\\
		\rightarrow&\frac{\eta_{BC}^N}{\sqrt{2}^{N-1}} |\text{GHZ}^{\pm}\>_b\otimes\Bigg[\anid_{1,H}...\anid_{N,H} + ... + \anid_{1,V}...\anid_{N,V}\Bigg]|vac\>,
		\label{cnt_bell_7}
	\end{aligned}
\end{equation}
where 
\begin{equation}
|{\rm GHZ}^\pm\rangle_b=\frac{1}{\sqrt{2}}\left(\anib_{1,D} ... \anib_{N,D} \pm \anib_{1,A} ... \anib_{N,A}\right)|vac\rangle.
\label{GHZ state}
\end{equation}
At this step, the heralding probability $P_{hr}^{(BC)}$ is equal to the summation of square of the coefficient multiplied by the $\prod_{i=1}^{N}\anid_i$ mode
%%%%%%%%%%%%%%%%%%%%%%
\begin{equation}
	P_{hr}^{BC} = \left(\frac{\eta_{BC}^N}{\sqrt{2}^{2N-1}}\right)^2 \times 2^N =  \frac{\eta_{BC}^{2N}}{2^{N-1}}.
\end{equation}
%%%%%%%%%%%%%%%%%%%%%%
The $2^N$ factor denotes the degeneracy of heralding signal. After the heralding detection, all $\hat{d}^{\dagger}_{i}$ modes are eliminated, and the remaining path modes represent a superposition of $\anib_{1,D}...\anib_{N,D}/\sqrt{2}$ and $\anib_{1,A}...\anib_{N,A}/\sqrt{2}$, which is a GHZ state. The relative phase of the GHZ state is determined by the combination of the photon detected in the heralding detectors. Since the phase of GHZ states can be adjusted through a feed-forward process based on heralding measurement outcomes, the success probability $P_{suc}^{(BC)}$ to obtain the GHZ state in this scheme becomes the square of the coefficient multiplied by the wave function in Eq.~\eqref{cnt_bell_7},	
\begin{equation}
	\begin{aligned}
		P_{suc}^{(BC)} = \frac{\eta_{BC}^{2N}}{2^{N-1}}.
		\label{cnt_bell_Pscc}
	\end{aligned}
\end{equation}

In BC scheme, heralding detection always indicates GHZ state and 
\begin{equation}
	H_{eff}^{(BC)}=\frac{P_{suc}^{(BC)}}{P_{hr}^{(BC)}}=1.
\end{equation}
There is correlation between the photons experiencing loss on their way to the heralding part and the photons being kept within the each party, the heralding signal guarantee entanglement regardless of loss. 

However, it is demanding to generate synchronized Bell states among distant parties. In the next section, we will relax this stringent condition of the initial state preparation by utilizing single-photons instead of entangled photon pairs.

\subsection{Single-photon inputs with centralized third-party (SC scheme)}\label{centr_single}

%%%%%%%%%%%%%%%%%%%%%%%%%%%%%%%%%%%%%%
\begin{figure}[h]
	\centering
	\includegraphics[width=0.75\textwidth]{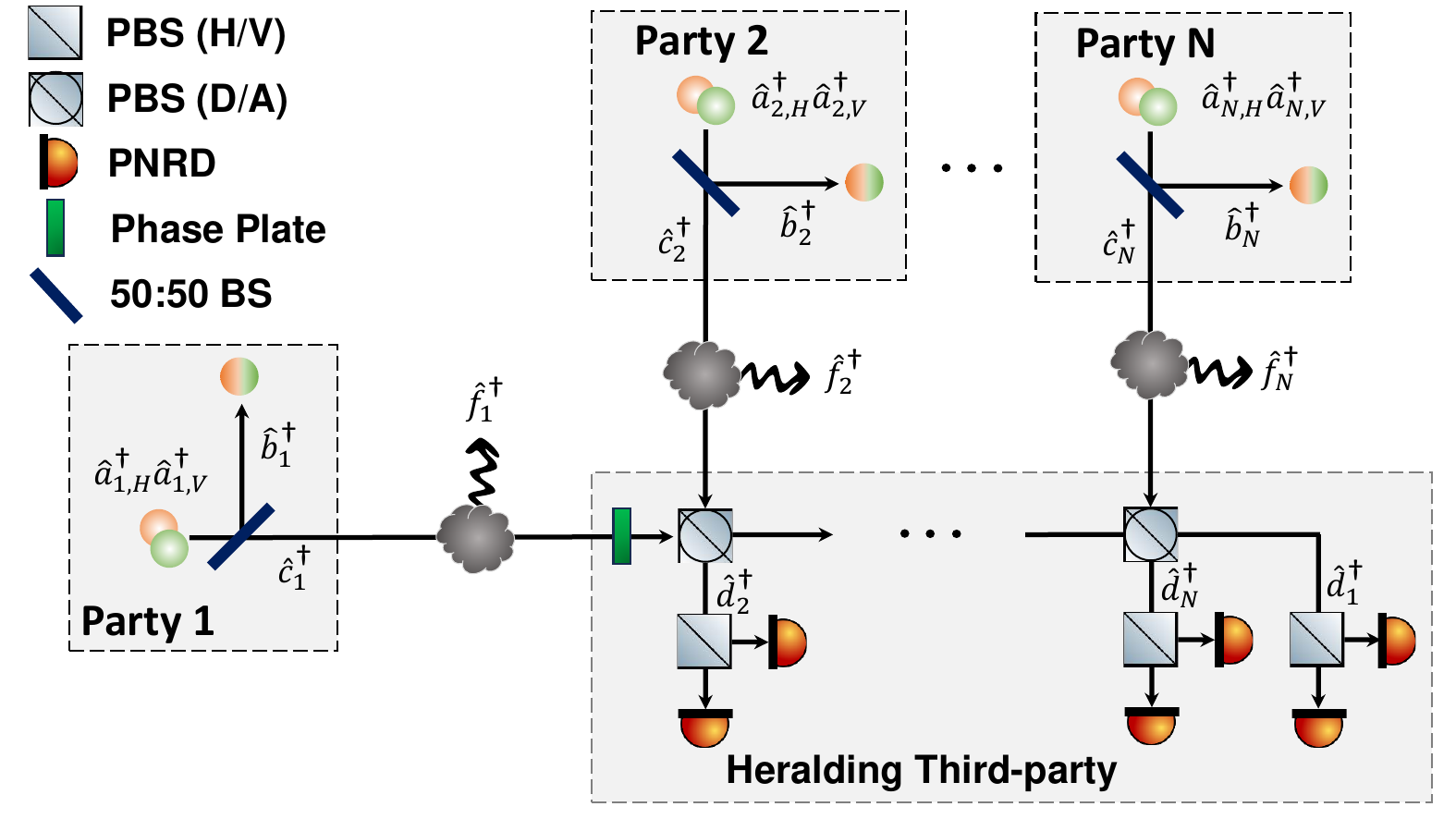}
	\caption{Schematic of the SC heralding scheme. Input resources are $2N$ single-photons, resulting in an $N$-party GHZ state.}
	\label{fig_central}
\end{figure}
%%%%%%%%%%%%%%%%%%%%%%%%%%%%%%%%%%%%%%
The centralized scheme with single-photon sources (SC scheme) is a generalization of the Bell state generation scheme suggested in Ref.~\cite{zhang2008,lasota2014linear}. This scheme shares similar network structure with the BC scheme but begins with $2N$ single-photon inputs instead of Bell states.

The initial state of the SC scheme is given by
\begin{equation}
	\begin{aligned}
		|\psi\>=\prod_{i=1}^{N}\ania_{i,H}\ania_{i,V}|vac\>.
		\label{cnt_1}
	\end{aligned}
\end{equation}
After passing through 50:50 non-polarizing beam splitter (BS) at each party, the path mode operators undergo following transformation: 		
\begin{equation}
	\begin{aligned}
		&\ania_{i,H}\rightarrow \frac{1}{\sqrt{2}}(\anib_{i,H}+\anic_{i,H}),~~~\ania_{i,V}\rightarrow \frac{1}{\sqrt{2}}(\anib_{i,V}+\anic_{i,V}).
		\label{cnt_2}
	\end{aligned}
\end{equation}
Then, the lossy channel with the transmission coefficient $\eta_{SC}$ transforms the mode operators,
\begin{equation}
	\anic_{i} \rightarrow \eta_{SC} \anic_{i} + \lossSC\anif_i,
\end{equation}
and the wavefunction becomes
\begin{equation}
	\begin{aligned}
		&|\psi\> \rightarrow \frac{1}{2^N} \prod_{i=1}^{N} \bigg[  \anib_{i,H}\anib_{i,V} + \eta_{SC} (\anib_{i,H}\anic_{i,V}+ \anib_{i,V} \anic_{i,H})+\eta_{SC}^2 \anic_{i,H} \anic_{i,V} \\
		& + \lossSC \left( \anib_{i,H}\anif_{i,V} + \anib_{i,V}\anif_{i,H}  +\eta_{SC}(\anic_{i,H}\anif_{i,V} +\anic_{i,V}\anif_{i,H})  \right)
		+(1-\eta_{SC}^2)\anif_{i,H}\anif_{i,V} \bigg]|vac\>.
		\label{cnt_3}
	\end{aligned}
\end{equation}	
As in the heralding part of Fig.~\ref{fig_central}, the path mode operators sent to the third-party transform into D/A basis same as Eq.~\eqref{cnt_bell_3}.

\begin{equation}
	\begin{aligned}
		\anic_{i,H}&\rightarrow& \frac{1}{\sqrt{2}} \left(\anid_{i,D} + \anid_{i\oplus 1,A} \right),~~~\anic_{i,V}&\rightarrow& \frac{1}{\sqrt{2}} \left(\anid_{i,D} - \anid_{i\oplus 1,A} \right)
		\label{cnt_33}
	\end{aligned}
\end{equation}

Then, the quantum state is:
\begin{equation}
	\begin{aligned}
		|\psi\> \rightarrow &~\frac{1}{2^N}\prod_{i=1}^N \bigg[ \anib_{i,H}\anib_{i,V}
		+ \frac{\eta_{SC}^2}{2} \left( \anid_{i,D}\anid_{i,D} - \anid_{i\oplus 1,A}\anid_{i\oplus 1,A} \right) \\
		&+ \frac{\eta_{SC}}{\sqrt{2}} \left( (\anib_{i,H}+\anib_{i,V})\anid_{i,D} + (\anib_{i,H}-\anib_{i,V})\anid_{i\oplus 1,A}\right) +\lossSC \left(\anib_{i,H}\anif_{i,V}+\anib_{i,V}\anif_{i,H}\right)\\
		&+\frac{\eta_{SC}\lossSC }{\sqrt{2}}\left( (\anid_{i,D}+\anid_{i\oplus 1,A})\anif_{i,V}+ (\anid_{i,D}-\anid_{i\oplus 1,A})\anif_{i,H} \right) 
		+\left( 1-\eta_{SC}^2 \right)\anif_{i,H}\anif_{i,V} \bigg]|vac\>.
		\label{cnt_4}
	\end{aligned}
\end{equation}
PNRD can postselect the cases that only one photon arrives at each heralding detectors. The terms that survive under the postselections are 
\begin{equation}
	\begin{aligned}
		&|\psi\> \rightarrow \frac{1}{2^N}\prod_{i=1}^N \Bigg[\eta_{SC} \left(\anib_{i,D}\anid_{i,D} + \anib_{i,A}\anid_{i\oplus 1,A}\right) \\ &+\frac{\eta_{SC}\lossSC }{\sqrt{2}} \left( (\anid_{i,D}+\anid_{i\oplus 1,A})\anif_{i,V}+ (\anid_{i,D}-\anid_{i\oplus 1,A})\anif_{i,H} \right) \Bigg]|vac\>,
	\end{aligned}
\end{equation}

and then,

\begin{equation}
	\begin{aligned}
		&|\psi\> \rightarrow \frac{\eta_{SC}^N}{2^N}  \Bigg[\prod_{i=1}^N \anib_{i,D}\anid_{i,D} + \prod_{j=1}^N \anib_{j,A}\anid_{j,A}\Bigg]|vac\> +\\
		& \frac{\eta_{SC}^{N-1}}{2^N} \left( \frac{\eta_{SC}\lossSC }{\sqrt{2}} \right) \anib_{1,D}\anib_{2,D}...\anib_{N-1,D}\anif_{N,V}\anid_{1,D}\anid_{2,D}...\anid_{N-1,D}\anid_{N,D}|vac\> \\
		& \frac{\eta_{SC}^{N-2}}{2^N} \left( \frac{\eta_{SC}\lossSC }{\sqrt{2}} \right)^2 \anib_{1,D}\anib_{2,D}...\anif_{N-1,V}\anif_{N,V}\anid_{1,D}\anid_{2,D}...\anid_{N-1,D}\anid_{N,D}|vac\>\\
		+...
		\label{cnt_psi_hr}
	\end{aligned}
\end{equation}

The first two terms in Eq.~\eqref{cnt_psi_hr} correspond to the generation of the GHZ state, see Eq.~\eqref{cnt_bell_6}. The remainders would cause false heralding signals that decreases the heralding efficiency, $H_{eff}$.

%However, when compared to the corresponding equation in the BC scheme (Eq.~\eqref{cnt_bell_6}), even after considering post-selection through PNRD, additional terms remain. 

\begin{figure}
	\centering
	\includegraphics[width=8.3cm]{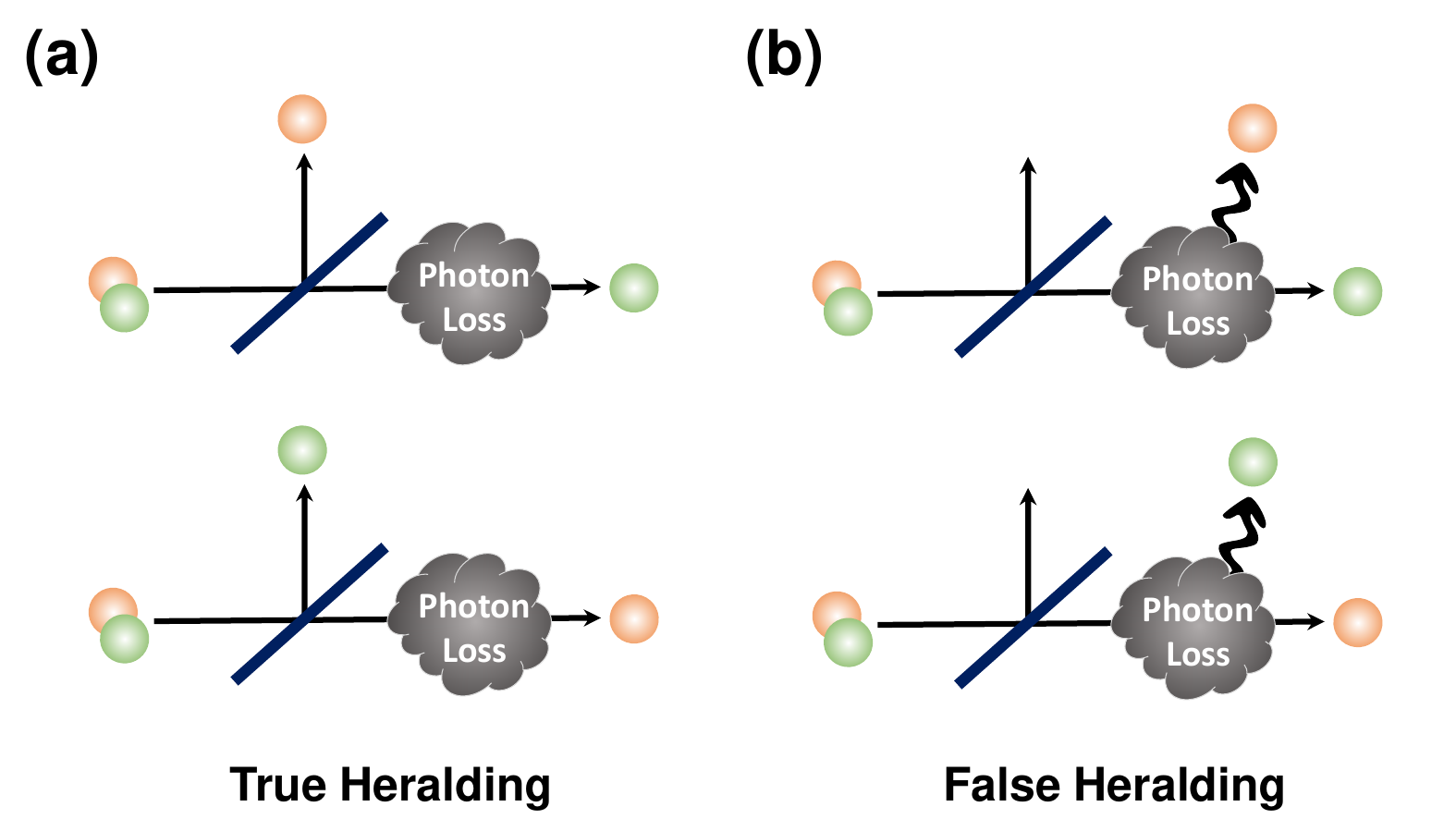}
	\caption{Scenarios leading to heralding signals in the SC scheme. (a) True heralding: One single-photon is retained by the local party, while the other successfully traverses the lossy channel to the central third-party. (b) False heralding: Both single-photons enter the lossy channel, but only one survives. The central heralding detector cannot distinguish this case from the true heralding scenario, leading to a false positive signal.}
	\label{fig_cnt_loss}
\end{figure}

Because the conditions on the inputs are relaxed, the correlation between the lost photon during the transmission and the photon kept by the local party for generating entanglement disappears. Our target is to generate an N-party GHZ state, where each party holds one photon and N photons are detected at the central location, as shown in Fig.~\ref{fig_cnt_loss}(a).
However, as illustrated in Fig.~\ref{fig_cnt_loss}(b), there are cases where photons arrive at the central location, but no photons remain at the local parties. Due to this difference, the heralding probability in the SC scheme is not equal to the success probability.

%%%%%%%%%%%%%%%%%%%%%%%%%%%%%%%%%%%%%%
%%%%%%%%%%%%%%%%%%%%%%%%%%%%%%%%%%%%%%

Therefore, taking into account both true heralding and false heralding cases, the heralding probability is calculated as follows:
%%%%%%%%%%%%%%%%%%%%%%%%%%%%%%%%%%%%%%
\begin{eqnarray}
	%\begin{aligned}
		P_{hr}^{(SC)} &=&\frac{1}{2^N}\bigg[ 2\eta_{SC}^{2N} +\sum_{1}^{N}  \binom{N}{k} (2\eta_{SC}^2(1-\eta_{SC}^2))^k (\eta_{SC}^2)^{N-k}\bigg] \nonumber\\
		&=& \frac{1}{2^N}\bigg[2\eta_{SC}^{2N}+(2\eta_{SC}^2(1-\eta_{SC}^2)^2+\eta_{SC}^2)^N -\eta_{SC}^{2N}\bigg].
		\label{cnt_PH}
	%\end{aligned}
\end{eqnarray}
%%%%%%%%%%%%%%%%%%%%%%%%%%%%%%%%%%%%%%
The success probability of entanglement distribution is obtained by taking the square of the amplitude of only the first two terms in Eq.~\eqref{cnt_psi_hr}, in the same manner as in Eq.~\eqref{cnt_bell_6} and Eq.~\eqref{cnt_bell_7}, and thus
%%%%%%%%%%%%%%%%%%%%%%%%%%%%%%%%%%%%%%
\begin{equation}
	\begin{aligned}
		P_{suc}^{(SC)} = \frac{\eta_{SC}^{2N}}{2^{2N-1}}.
		\label{cnt_Pscc}
	\end{aligned}
\end{equation}
%%%%%%%%%%%%%%%%%%%%%%%%%%%%%%%%%%%%%%
Then, we have the heralding efficiency of the SC scheme is given as
%%%%%%%%%%%%%%%%%%%%%%%%%%%%%%%%%%%%%%
\begin{equation}
	\begin{aligned}
		H_{eff}^{(SC)} = \frac{P_{suc}^{(SC)}}{P_{hr}^{(SC)}} =\frac{2}{1 + (3-2\eta_{SC}^2)^N}.
		\label{cnt_hr_eff}
	\end{aligned}
\end{equation}
%%%%%%%%%%%%%%%%%%%%%%%%%%%%%%%%%%%%%%

\subsection{Single-photon inputs with decentralized heralding (SD scheme)}\label{decentr_single}

The two schemes discussed so far rely on central measurements by a third-party that distributes the classical information on heralding detection results to $N$ participants. Hence, it raises potential trust concerns due to the information imbalance between the third-party and others~\cite{pramanik2020equitable}. Therefore, the generation of quantum entangled states with no trusted third-party are useful for enhancing  the security of quantum communications. We can consider such schemes based on the property that the photon path indistinguishability does not rely on direct photon-photon interactions but interference between probability amplitudes~\cite{krenn2017, kim2018, barros2020, sun2020, lee2022}. Indeed, Refs.~\cite{chin2024,chin2023} proposed a heralded scheme for generating the GHZ state without third-party via a graph approach to boson subtractions, which we call the decentralized scheme with single-photon sources (SD scheme). In this subsection, we calculate the success probability and heralding efficiency of the SD scheme in presence of channel losses.  

Initially, we have $N$ copies of one horizontal and one vertical polarization single-photons impinged into one port of a PBS, hence the initial state is given by
 %%%%%%%%%%%%%%%%%%%%%%%%%%%%%%%%%%%%%%
\begin{equation}
	\begin{aligned}
		|\psi\>=\prod_{i=1}^{N}\ania_{i,H}\ania_{i,V}|vac\rangle.
		\label{decen_init}
	\end{aligned}
\end{equation}
%%%%%%%%%%%%%%%%%%%%%%%%%%%%%%%%%%%%%%
The D/A basis PBS transform the photons as	
%%%%%%%%%%%%%%%%%%%%%%%%%%%%%%%%%%%%%%
\begin{equation}
	\begin{aligned}
		&\ania_{i,H}\rightarrow \frac{1}{\sqrt{2}}(\anib_{i,D}+\anic_{i,A}),~~~\ania_{i,V}\rightarrow \frac{1}{\sqrt{2}}(\anib_{i,D}-\anic_{i,A}).
		\label{decen_2}
	\end{aligned}
\end{equation}
%%%%%%%%%%%%%%%%%%%%%%%%%%%%%%%%%%%%%%
%%%%%%%%%%%%%%%%%%%%%%%%%%%%%%%%%%%%%%
\begin{figure}[t]
	\centering
	\includegraphics[width=0.75\textwidth]{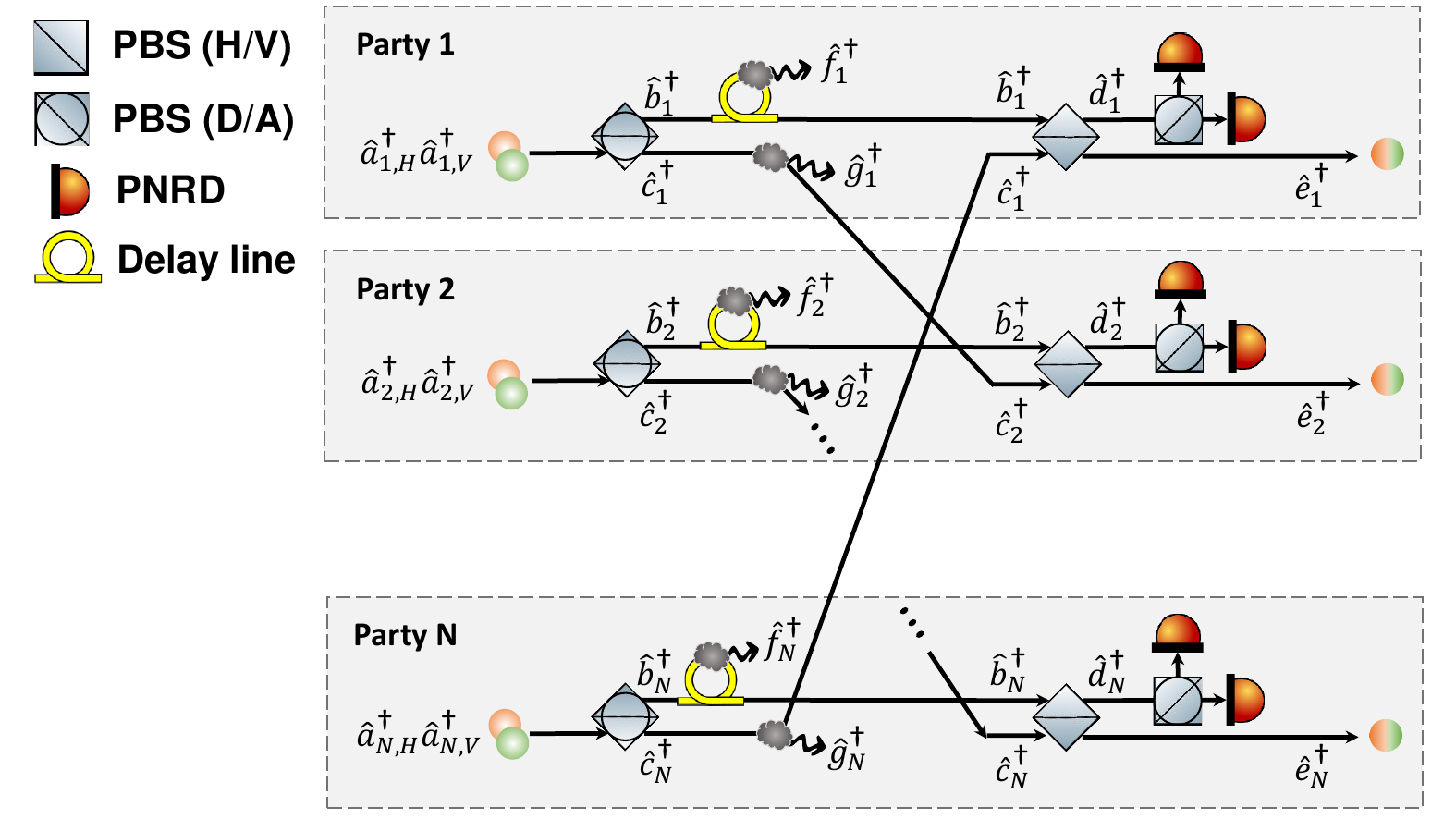}
	\caption{Schematic of the SD heralding scheme. Input resources are $2N$ single-photons, resulting in an $N$-party GHZ state. Heralding detection components are evenly distributed among all participants, eliminating the need for a centralized third party.}
	\label{fig_decen}
\end{figure}
%%%%%%%%%%%%%%%%%%%%%%%%%%%%%%%%%%%%%%
Hence the wave function becomes
%%%%%%%%%%%%%%%%%%%%%%%%%%%%%%%%%%%%%%
\begin{equation}
	\begin{aligned}
		&|\psi\> \rightarrow \frac{1}{2^N}\prod_{i=1}^{N} \left(\anib_{i,D}\anib_{i,D}-\anic_{i,A}\anic_{i,A}\right)|vac\>.
		\label{decen_hom}
	\end{aligned}
\end{equation} 
%%%%%%%%%%%%%%%%%%%%%%%%%%%%%%%%%%%%%%
Note that the Hong-Ou-Mandel (HOM) interference appears from the $N$ PBS, by which two single-photons always exit the same output port of a PBS~\cite{hong1987}.

After the HOM interference, the photons transmit to neighboring party or come back to their own party. During this process, photons can be lost and this is represented by 
%%%%%%%%%%%%%%%%%%%%%%%%%%%%%%%%%%%%%%
\begin{equation}
	\begin{aligned}
		&\anib_{i,D}\rightarrow \eta_{SD}\anib_{i,D} +\lossSD\anif_{i,D}, \\
		&\anic_{i,A}\rightarrow \eta_{SD} \anic_{i,A} +\lossSD\anig_{i,A}.
		\label{decen_loss}
	\end{aligned}
\end{equation}
%%%%%%%%%%%%%%%%%%%%%%%%%%%%%%%%%%%%%%
Then, the overall quantum state becomes
%%%%%%%%%%%%%%%%%%%%%%%%%%%%%%%%%%%%%%
\begin{equation}
	\begin{aligned}
		|\psi\> \rightarrow &\frac{1}{2^N}\prod_{i=1}^N\Bigg[ \eta_{SD}^2(\anib_{i,D}\anib_{i,D}-\anic_{i,A}\anic_{i,A})+2\eta_{SD}\lossSD\left(\anib_{i,D}\anif_{i,D}-\anic_{i,A}\anig_{i,A}\right) \\
		&+ \left(1-\eta_{SD}^2\right) \left(\anif_{i,D}\anif_{i,D}-\anig_{i,A}\anig_{i,A} \right) \Bigg]|vac\>.
	\end{aligned}
\end{equation}
%%%%%%%%%%%%%%%%%%%%%%%%%%%%%%%%%%%%%%

The optical path wiring is represented as 
%%%%%%%%%%%%%%%%%%%%%%%%%%%%%%%%%%%%%%
\begin{equation}
	\begin{aligned}
		\anib_{i,D} \rightarrow \anib_{i,D},~~~
		&\anic_{i,A} \rightarrow \anic_{i\oplus 1,A}.
		\label{decen_rewire} 
	\end{aligned}
\end{equation} 
%%%%%%%%%%%%%%%%%%%%%%%%%%%%%%%%%%%%%%

After the lossy channels and rewiring process, the photons arrive at the H/V basis PBS. 
%%%%%%%%%%%%%%%%%%%%%%%%%%%%%%%%%%%%%%
\begin{equation}
	\begin{aligned}
		\anib_{i,D} \rightarrow \frac{1}{\sqrt{2}}\left(\anie_{i,H} + \anid_{i,V}\right), ~~~\anic_{i,A} \rightarrow \frac{1}{\sqrt{2}}\left(\anid_{i,H} - \anie_{i,V}\right).
		\label{decen_HV} 
	\end{aligned}
\end{equation} 
%%%%%%%%%%%%%%%%%%%%%%%%%%%%%%%%%%%%%%
Now, the wave function is written in the H/V basis as
%%%%%%%%%%%%%%%%%%%%%%%%%%%%%%%%%%%%%%
\begin{equation}
	\begin{aligned}
		&|\psi\rangle \rightarrow \frac{1}{2^N} \prod_{i=1}^N 
		\Bigg[ \frac{\eta_{SD}^2}{2}\left( \anie_{i,H}\anie_{i,H}+\anid_{i,V}\anid_{i,V}+2\anie_{i,H}\anid_{i,V}
		-\anid_{i\oplus 1,H}\anid_{i\oplus 1,H}-2\anid_{i\oplus 1,H}\anie_{i\oplus 1,V}-\anie_{i\oplus 1,V}\anie_{i\oplus 1,V} \right)\\
		 &+\sqrt{2}\eta_{SD}\lossSD\left( \left( \anie_{i,H}+\anid_{i,V} \right)\anif_{i,D} -\left( \anid_{i\oplus 1,H} -\anie_{i\oplus 1,V} \right)\anig_{i,A} \right)
		 +(1-\eta_{SD}^2)\left(\anif_{i,D}\anif_{i,D}-\anie_{i,V}\anie_{i,V}\right)
		 \Bigg]|vac\>.
		\label{decen_7}
	\end{aligned}
\end{equation}
%%%%%%%%%%%%%%%%%%%%%%%%%%%%%%%%%%%%%%
The heralding probability $P_{hr}^{(SD)}$ is derived from the combination of terms, each containing only one detector path mode $\anid$ as 
%%%%%%%%%%%%%%%%%%%%%%%%%%%%%%%%%%%%%%
\begin{equation}
	\begin{aligned}
		&|\psi\rangle \rightarrow \frac{1}{2^N} \prod_{i=1}^N 
		\Bigg[ \eta_{SD}^2\left(\anie_{i,H}\anid_{i,V}
		-\anid_{i\oplus 1,H}\anie_{i\oplus 1,V}\right)+\sqrt{2}\eta_{SD}\lossSD\left(\anid_{i,V}\anif_{i,D} - \anid_{i\oplus 1,H}\anig_{i,A} \right)
		\Bigg]|vac\>.
		\label{decen_hr}
	\end{aligned}
\end{equation}
%%%%%%%%%%%%%%%%%%%%%%%%%%%%%%%%%%%%%%
%Specifically, these terms are  $\eta_{SD}^2 \anid_{i,H}\anie_{i,V}$, $\eta_{SD}^2 \anid_{i\oplus 1,V}\anie_{i\oplus 1,H}$, and $\sqrt{2}\eta_{SD}\lossSD \anid_{i\oplus 1 , V}\anif_{i,D}$. 
The heralding probability is calculated by taking the product of these terms and becomes
%%%%%%%%%%%%%%%%%%%%%%%%%%%%%%%%%%%%%%
\begin{equation}
		P_{hr}^{(SD)} =\frac{1}{2^{2N}}\bigg[ 2\eta_{SD}^{4N} +\sum_{1}^{N}   \binom{N}{k}  (2\eta_{SD}^2(1-\eta_{SD}^2))^k (\eta_{SD}^4)^{N-k}\bigg] = \frac{1}{2^{2N}}\bigg[ (2\eta_{SD}^2-\eta_{SD}^4)^N +\eta_{SD}^{4N}  \bigg].
		\label{decen_PH}
\end{equation}
%%%%%%%%%%%%%%%%%%%%%%%%%%%%%%%%%%%%%%

To distribute a GHZ state, out of $2N$ photons, $N$ must be detected by the heralding detectors, leaving $N$ photons. The quantum state and the success probability $P_{suc}^{SD}$ with feed-forward are as follows:
%%%%%%%%%%%%%%%%%%%%%%%%%%%%%%%%%%%%%%
\begin{equation}
	\begin{aligned}
		&|\psi\> \rightarrow \frac{\eta_{SD}^{2N}}{2^N} \bigg[ \prod_{i=1}^N  \hat{d}^{\dagger}_{i,H}\hat{e}^{\dagger}_{i,V} + \prod_{i=1}^N \hat{d}^{\dagger}_{i,V}\hat{e}^{\dagger}_{i,H} \bigg]|vac\rangle,
		\label{decen_8}
	\end{aligned}
\end{equation}
%%%%%%%%%%%%%%%%%%%%%%%%%%%%%%%%%%%%%%
\begin{equation}
	\begin{aligned}
		P_{suc}^{(SD)} = \frac{\eta_{SD}^{4N}}{2^{2N-1}}.
		\label{decen_Pscc}
	\end{aligned}
\end{equation}
%%%%%%%%%%%%%%%%%%%%%%%%%%%%%%%%%%%%%%
Then, we have the heralding efficiency of the SD scheme as
%%%%%%%%%%%%%%%%%%%%%%%%%%%%%%%%%%%%%%
\begin{equation}
	\begin{aligned}
		H_{eff}^{(SD)} = \frac{P_{suc}^{(SD)}}{P_{hr}^{(SD)}} = \frac{2\eta_{SD}^{2N}}{(2-\eta_{SD}^2)^N +\eta_{SD}^{2N} }.
		\label{decen_hr_eff}
	\end{aligned}
\end{equation}
%%%%%%%%%%%%%%%%%%%%%%%%%%%%%%%%%%%%%%

\section{Success probability and heralding efficiency of distribution structures}\label{comparative_study}
	
In this section, we evaluate the three heralded entanglement distribution schemes based on the criteria of network structures, success probabilities, and heralding efficiencies in the presence of channel losses. Nowadays, with the proposal and implementation of inter-city or metropolitan quantum networks, it is of importance to investigate the performances based on realistic network structure and loss values~\cite{azuma2015all,dynes2019cambridge,chen2021implementation,chen2021TF-QKD,chung2022,shen2023}. This analysis will uncover the relative advantages of each scheme over the others from various perspectives. 
		
		We can arrange the $N$ parties in the three heralded schemes so that they interchange quantum and classical information through the most efficient channels, which is the two-dimensional circular structure as drawn in Fig.~\ref{scheme}. We denote the radius of the circle as $R$ (km).

In order to consider the practical quantum network scenario via optical fibers, we assume that the channel transmission coefficient $\eta$ decreases exponentially along the channel length, i.e.,
		\begin{align}
			\eta (l) = e^{-\alpha l},
		\end{align} 
		where $\alpha$ is the attenuation constant considering an optical fiber channel loss and $l$ the optical channel length. Specifically, nowadays fiber loss is approximately 4.45\,\% per kilometer, thus we assume that $\alpha = 0.023~{\rm km^{-1}}$ corresponding to $|\eta\, (1\,{\rm km})|^2\simeq 0.955$ at the optical communication wavelength\cite{sangouard2011quantum}. Note that, we presume that $\eta$ is a real number because the relative phase can be compensated with up-to-date technologies over hundreds of kilometers in quantum communication~\cite{wang2022twin,liu2023TF}.
	
	Table~\ref{table_compare} summarizes the essential properties of the three heralded schemes. The BC scheme has a higher success probability than the others and its heralding efficiency is always $H_{eff}=1$ irrespective of the channel transmission $\eta_{BC}$. However, it requires  synchronized and deterministic $N$ Bell state sources to be connected with the centralized structure. Therefore, the SC and SD schemes that exploit single-photon sources have a definite advantage over the BC scheme. As explained in Sec.~\ref{decentr_single}, informationally balanced multi-party quantum communication without a trusted third party is possible with the SD scheme. However, from the viewpoint of the success probability and heralding efficiency, our analysis shows that the SC scheme can be better than the SD scheme according to the number or parties $N$ and the network circle radius $R$.  
	
	%%%%%%%%%%%%%%%%%%%%%%%%%%%%%%%%%%%%%%
  	\bgroup
		\def\arraystretch{2}%  1 is the default, change whatever
		\begin{table}[t]
			\centering
			\resizebox{0.85\textwidth}{!}{%
				\begin{tabular}{|c|c|c|c|}
					\hline
					&  BC scheme         &  SC scheme              &  SD scheme                 \\ \hline
					Photon sources & ~~Bell states~~  & Single-photons   & Single-photons      \\ \hline
					Network structure     & Centralized & Centralized      & Decentralized     \\ \hline
					Success probability ($P_{suc}$)    & $\frac{(\eta_{BC})^{2N}}{2^{N-1}}$ &  $\frac{(\eta_{SC})^{2N}}{2^{2N-1}}$ & $\frac{(\eta_{SD})^{4N}}{2^{2N-1}}$       \\ \hline
					Heralding efficiency ($H_{eff}$)  & $1$ &  $\frac{2}{1 + (3-2(\eta_{SC})^2)^N}$     & $\frac{2(\eta_{SD})^{2N}}{(2-(\eta_{SD})^2)^N +\eta_{SD}^{2N} }$       \\ \hline						
				\end{tabular}
			}
			\caption{Comparison of three heralded entanglement distribution schemes. In the two-dimensional round network geometry, the effective channel lengths $l$ can be $l_{BC}=l_{SC}=R$ and $l_{SD}=2R\sin{\pi/N}$ where $R$ is the radius of circle, and thus, $\eta_{BC}=\eta_{SC}= e^{-\alpha R}$ and $\eta_{SD}= e^{-2\alpha R\sin{\pi/N}}$, respectively. See the network geometry in Fig.~\ref{scheme}.}
			\label{table_compare}
		\end{table}
		\egroup
%%%%%%%%%%%%%%%%%%%%%%%%%%%%%%%%%%%%%%
	
\begin{figure}[t]
		\centering
		\includegraphics[width=\textwidth]{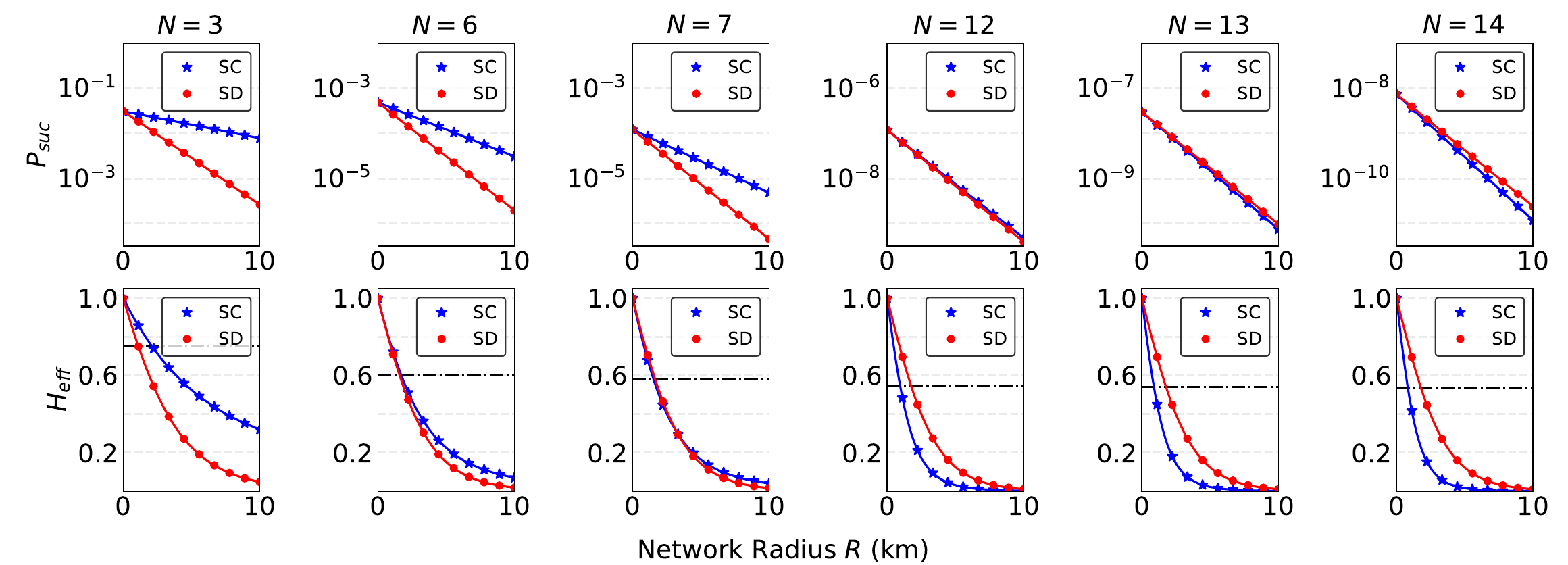}
		\caption{Comparison of success probability ($P_{suc}$) and heralding efficiency ($H_{eff}$) for the SC and SD schemes. Top row: Success probabilities, demonstrating $P_{suc}^{(SC)} > P_{suc}^{(SD)}$ for $N \leq 12$. Bottom row: Heralding efficiencies, showing the existence of non-zero cross-over radii for $N\geq 7$, where $H_{eff}^{(SC)} = H_{eff}^{(SD)}$. The dashed black horizontal lines are the threshold heralding efficiency for a local hidden variable model $H_{th}=\frac{N}{2N-2}$~\cite{cabello08}.}
		\label{fig_graph}
	\end{figure}

We first consider the success probabilities of the schemes, see the first row of graphs in Fig.~\ref{fig_graph}(a). We should note that the channel lengths $l$ among subsystems for the SC and SD schemes are $l_{SC}=R$ and $l_{SD}=2R\sin{(\pi/N)} $, respectively, see Fig.~\ref{scheme}, (c) and (d). Hence the channel length for the SD scheme becomes smaller than the SC scheme as $N$ increases. On the other hand, since the SD scheme sends two photons in each lossy channel while the SC scheme sends one photon, the success probability of the SD scheme is more sensitive to the channel loss than the SC scheme. Based on these observations, we expect that there exist a certain number $N$ that the success probability of the SD scheme becomes larger than that of the SC scheme. Indeed, we can directly check from the third row of Table~\ref{table_compare} that the following inequality always holds
	\begin{align}
		P_{suc}^{(SC)} > P_{suc}^{(SD)}, ~~~~~{\rm for}~~ N \le 12,\\
		P_{suc}^{(SC)} < P_{suc}^{(SD)}, ~~~~~{\rm for}~~ N > 12.\nonumber
		\end{align}

	% : Check if this sentence is correct. If this is true, we can present two probabilities with the same equation number. Also let us add $N=14, 18$ or other larger $N$ in Fig. 6} %\teal{(Wan: I added N=14 graph. Is there any specific reason for N=18? If there are too many graphs, it becomes difficult to see. N=12-14 would be sufficient to explain the crossover.)}
	
	%%%%%%%%%%%%%%%%%%%%%%%%%%%
	%%%%%%%%%%%%%%%%%%%%%%%%%%%%%%%%%%%%%%
	
	%%%%%%%%%%%%%%%%%%%%%%%%%%%
	\begin{figure}[t]
		\centering
		\includegraphics[width=0.5\textwidth]{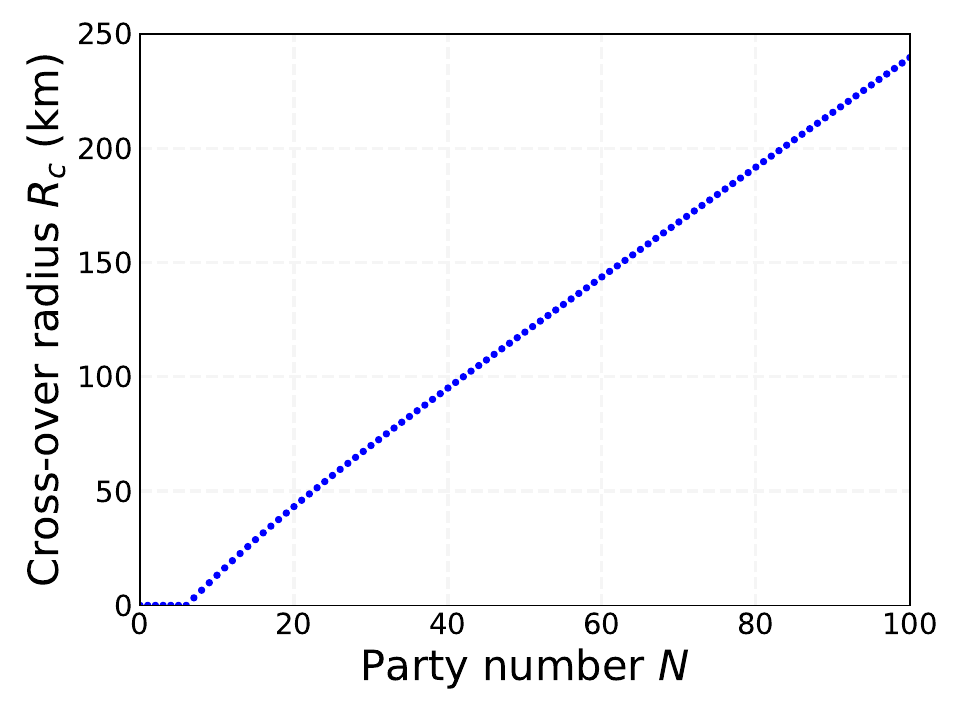}
		\caption{Cross-over radius $R_c$ as a function of party number $N$. $R_c$ remains zero for $N \leq 6$ and increases monotonically for $N \geq 7$. The near-linear relationship between $R_c$ and $N$ for large $N$ implies that the inter-party distance $l_c$ at which $H_{eff}^{(SD)}=H_{eff}^{(SC)}$ becomes nearly independent of $N$ for large networks.}
		\label{fig_Lc}
	\end{figure}
	%%%%%%%%%%%%%%%%%%%%%%%%%%%%%%%%%%%%%%
	
Next, we can find a more interesting pattern in the behaviors of the heralding efficiencies as depicted at the second row of Fig.~\ref{fig_graph}. For comparison, we present the threshold heralding efficiency for a local hidden variable model $H_{th}=\frac{N}{2N-2}$ as the horizontal dashed black lines~\cite{cabello08}. From the $H_{eff}^{(SC)}$ and $H_{eff}^{(SD)}$ given by Eqs.~\eqref{cnt_hr_eff} and \eqref{decen_hr_eff}, we can see that a cross-over radius $R_c$ exists at which $H_{eff}^{(SD)}$ is equal to $H_{eff}^{(SC)}$, which satisfies the following equality
	\begin{equation}
		\begin{aligned}
			e^{-2\alpha R_c}+e^{4\alpha R_c sin(\pi/N)}=2.
			\label{crossover}
		\end{aligned}
	\end{equation}
This gives $R_c=0$ for $N\le 6$. This means that $H_{eff}^{(SC)}>H_{eff}^{(SD)}$ in any radii for those cases (see the first two graphs of the second row in Fig.~\ref{fig_graph}). For $N\ge 7$, $R_c$ becomes nonzero (see the other graphs of the second raw in Fig.~\ref{fig_graph}), which implies that $H_{eff}^{(SC)}<H_{eff}^{(SD)}$ for $R$ that is smaller than $R_c$. 

We can directly check that $R_c$ increases monotonically as $N$ increases, as shown in Fig.~\ref{fig_Lc}. This implies that the distance between adjacent parties at which $H_{eff}^{(SD)}$ is equal to $H_{eff}^{(SC)}$ (which is expressed as $l_c=2R_c\sin(\pi /N)$) is almost independent of $N$ for the case. We can easily verify that $l_c$ converges to $\sim 15.71$ km as $N$ increases since $\sin(1/N)\simeq 1/N$ for large $N$. Therefore, the heralding efficiency of the SD scheme is larger than that of SC scheme for large size of network ($N\ge7$) when each party is closer than $\sim 15.71~km$ from its adjacent parties.

%Therefore, we can also state that \emph{the SC scheme is more useful than the SD scheme for a large size of network, for \teal{$H_{eff}^{(SC)}>H_{eff}^{(SD)}$} always holds when each party is farther than $\sim 15.71~km$ from its adjacent parties.}
  
To summarize, we have shown the following relations between the SC and SD schemes:
\begin{enumerate}
    \item 	$P_{suc}^{(SC)} > P_{suc}^{(SD)}$ for $N\le 12$ and $P_{suc}^{(SC)} < P_{suc}^{(SD)}$ for $N> 12$.
    \item $H_{eff}^{(SC)} > H_{eff}^{(SD)}$ in any radii for $N \le 6$.
    \item $H_{eff}^{(SC)} < H_{eff}^{(SD)}$ when the inter-party distance is smaller than $\sim 15.71$~km for any $N$.
    \item The SD scheme offers information balance among the parties due to the symmetric network structure without a third party.
\end{enumerate}

	\section{Conclusion}\label{conclusion}
	
In this work, we presented a comparative study of three heralded schemes for distributing multipartite GHZ states over lossy quantum channels: the Bell state inputs with centralized third-party (BC) scheme, single-photon inputs with centralized third-party (SC) scheme, and single-photon inputs with decentralized heralding (SD) scheme. Our analysis focused on their success probabilities and heralding efficiencies in the presence of channel losses. We found that while the BC scheme demonstrated the highest success probability and constant heralding efficiency, its requirement for synchronized Bell state sources presents practical challenges. The SC and SD schemes, utilizing single-photon sources, can offer more practical alternatives.

Our results revealed that the SC scheme outperforms the SD scheme in terms of success probability $P_{suc}$ for networks with 12 or fewer parties. We also identified a cross-over radius $R_c$ in heralding efficiency between the SC and SD schemes for networks with 7 or more parties. Notably, for large networks, the SD scheme maintains superior heralding efficiency when the inter-party distance is smaller than approximately 15.71~km, regardless of the total number of parties. We also note that the SD scheme offers an important advantage in terms of information balance among parties. Unlike the SC scheme, which relies on a centralized third party for measurements, the SD scheme distributes heralding detectors evenly among all participants, addressing potential trust concerns arising from information imbalance in centralized schemes.

These results provide crucial insights for designing quantum networks, offering a framework for selecting the most appropriate entanglement distribution scheme based on network size, inter-party distances, and security requirements. Future research could focus on investigating the impact of imperfect sources and detectors, exploring hybrid approaches, and analyzing the effects of decoherence in realistic quantum channels. %Our findings contribute significantly to the ongoing development of efficient and secure quantum communication infrastructures.

	\section*{Acknowledgements}
	\noindent This research was funded by the KIST institutional project (2E32941) and IITP-MSIP (RS-2023-00222863, RS-2024-00396999). SC is supported by National Research Foundation of Korea (NRF, RS-2023-00245747).
	
		\section*{Disclosures}
	\noindent The authors declare no conflicts of interest.

	%IITP-MSIP ??? QKD (RS-2024-00396999)
	%%%%%%%%%% If using BibTeX:
%\bibliography{sample} 

\begin{thebibliography}{99}

%Quantum networks
\bibitem{ursin2007entanglement} R. Ursin, {\it et al}., ``Entanglement-based quantum communication over 144 km,'' Nature Phys. {\bf 3}, 481--486 (2007).
		
\bibitem{simon2017towards} C. Simon, ``Towards a global quantum network,'' Nature Photon. {\bf 11}, 678--680 (2017).
		
\bibitem{wehner2018} S. Wehner, D. Elkouss, and R. Hanson, ``Quantum internet: A vision for the road ahead,'' Science {\bf 362}, eaam9288 (2018).
		
%Quantum cryptography with entanglement
\bibitem{ekert1991quantum} A. K. Ekert, ``Quantum cryptography based on Bell's theorem,'' Phys. Rev. Lett. {\bf 67}, 661 (1991).
		
%\bibitem{jennewein2000quantum} T. Jennewein, C. Simon, G. Weihs, H. Weinfurter, and A. Zeilinger, ``Quantum cryptography with entangled photons,'' Phys. Rev. Lett. 84, 4729 (2000).
		
\bibitem{yin2020entanglement} J. Yin, {\it et al}., ``Entanglement-based secure quantum cryptography over 1,120 kilometres,'' Nature {\bf 582}, 501--505 (2020).

%Distributed quantum sensing
\bibitem{liu2021distributed} L.-Z. Liu, {\it et al}., ``Distributed quantum phase estimation with entangled photons,'' Nature Photon. {\bf 15}, 137--142 (2021).
		
\bibitem{kim2024distributed} D.-H. Kim, S. Hong, Y.-S. Kim, Y. Kim, S.-W. Lee, R. C. Pooser, K. Oh, S.-Y. Lee, C. Lee, and H.-T. Lim, ``Distributed quantum sensing of multiple phases with fewer photons,'' Nature Commun. {\bf 15}, 266 (2024).

% Quantum computing
\bibitem{klm2001} E. Knill, R. Laflamme, and G. J. Milburn, ``A scheme for efficient quantum computation with linear optics,'' Nature {\bf 409}, 46--52 (2001).
		
\bibitem{barz2012demonstration} S. Barz, E. Kashefi, A. Broadbent, J. F. Fitzsimons, A. Zeilinger, and P. Walther, ``Demonstration of blind quantum computing,'' Science {\bf 335}, 303--308 (2012).

\bibitem{fitzsimons2017private} J. F. Fitzsimons, ``Private quantum computation: an introduction to blind quantum computing and related protocols,'' npj Quantum Inf. {\bf 3}, 23 (2017).

\bibitem{huang2017experimental} H.-L. Huang, {\it et al}., ``Experimental blind quantum computing for a classical client,'' Phys. Rev. Lett. {\bf 119}, 050503 (2017).

%QND measurements
				
\bibitem{braginsky1996quantum} V. B. Braginsky and F. Ya. Khalili, ``Quantum nondemolition measurements: the route from toys to tools,'' Rev. Mod. Phys. {\bf 68}, 1 (1996).
		
\bibitem{grangier1998quantum} P. Grangier, J. A. Levenson, and J.-P. Poizat, ``Quantum non-demolition measurements in optics,'' Nature {\bf 396}, 537--542 (1998).


%Heralded entanglement generation
\bibitem{walther2007heralded} P. Walther, M. Aspelmeyer, and A. Zeilinger, ``Heralded generation of multiphoton entanglement,'' Phys. Rev. A {\bf 75}, 012313 (2007).
		
\bibitem{wagenknecht2010} C. Wagenknecht, C.-M. Li, A. Reingruber, X.-H. Bao, A. Goebel, Y.-A. Chen, Q. Zhang, K. Chen, and J.-W. Pan, ``Experimental demonstration of a heralded entanglement source,'' Nature Photon. {\bf 4}, 549--552 (2010).
		
\bibitem{caspar2020heralded} P. Caspar, {\it et al}., ``Heralded distribution of single-photon path entanglement,'' Phys. Rev. Lett. {\bf 125}, 110506 (2020).
		
\bibitem{gubarev2020improved} F. V. Gubarev, I. V. Dyakonov, M. Yu. Saygin, G. I. Struchalin, S. S. Straupe, and S. P. Kulik, ``Improved heralded schemes to generate entangled states from single photons,'' Phys. Rev. A {\bf 102}, 012604 (2020).

% Entanglement swapping
		
\bibitem{pan1998experimental} J.-W. Pan, D. Bouwmeester, H. Weinfurter, and A. Zeilinger, ``Experimental entanglement swapping: entangling photons that never interacted,'' Phys. Rev. Lett. {\bf 80}, 3891 (1998).

\bibitem{pan1998greenberger} J.-W. Pan and A. Zeilinger, ``Greenberger-Horne-Zeilinger-state analyzer,'' Phys. Rev. A 57, 2208 (1998).

\bibitem{bose1998multiparticle} S. Bose, V. Vedral, and P. L. Knight, ``Multiparticle generalization of entanglement swapping,'' Phys. Rev. A 57, 822 (1998).

%GHZ state generation

\bibitem{bouwmeester1999} D. Bouwmeester, J.-W. Pan, M. Daniell, H. Weinfurter, and A. Zeilinger, ``Observation of three-photon Greenberger-Horne-Zeilinger entanglement,'' Phys. Rev. Lett. {\bf 82}, 1345 (1999).

\bibitem{brunner2014bell} N. Brunner, D. Cavalcanti, S. Pironio, V. Scarani, and S. Wehner, "Bell nonlocality," Rev. Mod. Phys. 86, 419--478 (2014).

\bibitem{li2015resource} Y. Li, P. C. Humphreys, G. J. Mendoza, and S. C. Benjamin, "Resource costs for fault-tolerant linear optical quantum computing," Phys. Rev. X 5, 041007 (2015).



%Heralded entanglement distribution with single-photon inputs

\bibitem{zhang2008} Q. Zhang, X.-H. Bao, C.-Y. Lu, X.-Q. Zhou, T. Yang, T. Rudolph, and J.-W. Pan, ``Demonstration of a scheme for the generation of `event-ready' entangled photon pairs from a single-photon source,'' Phys. Rev. A {\bf 77}, 062316 (2008).
		
\bibitem{lasota2014linear} M. Lasota, C. Radzewicz, K. Banaszek, and R. Thew, ``Linear optics schemes for entanglement distribution with realistic single-photon sources,'' Phys. Rev. A {\bf 90}, 033836 (2014).

\bibitem{chin2023} S. Chin, M. Karczewski, and Y.-S. Kim, ``Heralded Optical Entanglement Generation via the Graph Picture of Linear Quantum Networks." arXiv preprint arXiv:2310.10291 (2023).

\bibitem{chin2024} S. Chin, Y.-S. Kim, and M. Karczewski, ``Shortcut to multipartite entanglement generation: A graph approach to boson subtractions,'' npj Quantum Inf. {\bf 10}, 67 (2024).

\bibitem{pramanik2020equitable} T. Pramanik, D.-H. Lee, Y.-W. Cho, H.-T. Lim, S.-W. Han, H. Jung, S. Moon, K. J. Lee, and Y.-S. Kim, ``Equitable multiparty quantum communication without a trusted third party,'' Phys. Rev. Appl. {\bf 14}, 064074 (2020).

% spatial overlap

\bibitem{krenn2017} M. Krenn, A. Hochrainer, M. Lahiri, and A. Zeilinger, ``Entanglement by Path Identity,'' Phys. Rev. Lett. {\bf 118}, 080401 (2017).


\bibitem{kim2018} Y.-S. Kim, T. Pramanik, Y.-W. Cho, M. Yang, S.-W. Han, S.-Y. Lee, M.-S. Kang, and S. Moon, ``Informationally symmetrical Bell state preparation and measurement,'' Opt. Express {\bf 26}, 29539--29549 (2018).

\bibitem{barros2020} M. R. Barros, S. Chin, T. Pramanik, H.-T. Lim, Y.-W. Cho, J. Huh, and Y.-S. Kim, ``Entangling bosons through particle indistinguishability and spatial overlap,'' Opt. Express {\bf 28}, 38083--38092 (2020).

\bibitem{sun2020} K. Sun, {\it et al}., ``Experimental quantum entanglement and teleportation by tuning remote spatial indistinguishability of independent photons,'' Opt. Lett. {\bf 45}, 6410--6413 (2020).

\bibitem{lee2022} D. Lee, T. Pramanik, S. Hong, Y.-W. Cho, H.-T. Lim, S. Chin, and Y.-S. Kim, ``Entangling three identical particles via spatial overlap,'' Opt. Express {\bf 30}, 30525--30535 (2022).


\bibitem{hong1987} C. K. Hong, Z. Y. Ou, and L. Mandel, ``Measurement of subpicosecond time intervals between two photons by interference,'' Phys. Rev. Lett. {\bf 59}, 2044--2046 (1987).


% Metropolitan quantum network

%\bibitem{chen2010} T.-Y. Chen, J. Wang, H. Liang, W.-Y. Liu, Y. Liu, X. Jiang, Y. Wang, X. Wan, W.-Q. Cai, L. Ju, L.-K. Chen, L.-J. Wang, Y. Gao, K. Chen, C.-Z. Peng, Z.-B. Chen, and J.-W. Pan, ``Metropolitan all-pass and inter-city quantum communication network,'' Opt. Express 18, 27217--27225 (2010).
		
\bibitem{azuma2015all} K. Azuma, K. Tamaki, and W. J. Munro, ``All-photonic intercity quantum key distribution,'' Nat. Commun. {\bf 6}, 10171 (2015).
		
\bibitem{dynes2019cambridge} J. F. Dynes, {\it et al}., ``Cambridge quantum network,'' npj Quantum Inf. {\bf 5}, 101 (2019).
		
\bibitem{chen2021implementation} T.-Y. Chen, {\it et al}., ``Implementation of a 46-node quantum metropolitan area network,'' npj Quantum Inf. {\bf 7}, 134 (2021).	

\bibitem{chen2021TF-QKD} J.-P. Chen, {\it et al}., ``Twin-field quantum key distribution over a 511~km optical fibre linking two distant metropolitan areas,'' Nature Photon. {\bf 15}, 570--575 (2021).

\bibitem{chung2022} J. Chung, {\it et al}., ``Design and implementation of the Illinois Express quantum metropolitan area network,'' IEEE Trans. Quantum Eng. {\bf 3}, 1--20 (2022)

\bibitem{shen2023} S. Shen, {\it et al}., ``Hertz-rate metropolitan quantum teleportation,'' Light Sci. Appl. {\bf 12}, 115 (2023).

\bibitem{sangouard2011quantum} N. Sangouard, C. Simon, H. de Riedmatten, and N. Gisin, ``Quantum repeaters based on atomic ensembles and linear optics,'' Rev. Mod. Phys. {\bf 83}, 33--80 (2011).

\bibitem{wang2022twin} S. Wang, {\it et al}., ``Twin-field quantum key distribution over 830-km fibre,'' Nature Photon. {\bf 16}, 154--161 (2022).

\bibitem{liu2023TF} Y. Liu, {\it et al}., ``Experimental Twin-Field Quantum Key Distribution over 1000 km Fiber Distance,'' Phys. Rev. Lett. {\bf 130}, 210801 (2023).
		
\bibitem{cabello08} A. Cabello, D. Rodriguez, and I. Villanueva, ``Necessary and Sufficient Detection Efficiency for the Mermin Inequalities,'' Phys. Rev. Lett. {\bf 101}, 120402 (2008).  
		
\end{thebibliography}
%%%%%%%%%% If preparing manually:

\end{document}